**Defects and Phase Formation in Non-Stoichiometric LaFeO$_3$: a Combined Theoretical and Experimental Study**

Daniel Mutter[1]*, Roland Schierholz[2]*, Daniel F. Urban[1], Sabrina A. Heuer[2,3], Thorsten Ohlerth[2,3], Hans Kungl[2], Christian Elsässer[1,4], and Rüdiger-A. Eichel[2,3]

[1]Fraunhofer IWM, Wöhlerstraße 11, 79108 Freiburg, Germany

[2]Forschungszentrum Jülich GmbH, Institute of Energy and Climate Research (IEK-9), Wilhelm-Johnen-Straße, 52425 Jülich, Germany

[3]RWTH Aachen University, Institute of Physical Chemistry, Landoltweg 2, 52074 Aachen, Germany

[4]Freiburg Materials Research Center (FMF), University of Freiburg, Stefan-Meier-Straße 21, 79104 Freiburg, Germany

*corresponding author: daniel.mutter@iwm.fraunhofer.de

*corresponding author: r.schierholz@fz-juelich.de

**Abstract**

The defect chemistry of perovskite compounds is directly related to the stoichiometry and to the valence states of the transition-metal ions. Defect engineering has become increasingly popular as it offers the possibility to influence the catalytic properties of perovskites for applications in energy storage and conversion devices such as solid-oxide fuel- and electrolyzer cells. LaFeO$_3$ (LFO) can be regarded as a base compound of the family of catalytically active perovskites La$_{1-x}$A$_x$Fe$_{1-y}$B$_y$O$_{3-\delta}$, for which the defect chemistry as well as the electronic and ionic conductivity can be tuned by substitution on cationic sites. Combining theoretical and experimental approaches, we explore the suitability for A-site vacancy engineering, namely the feasibility of actively manipulating the valence state of Fe and the concentration of point defects by synthesizing La-deficient LFO. In the theoretical part, formation energies and concentrations of point defects were determined as a function of processing conditions by first-principles density functional theory calculations with a Hubbard-$U$ correction (DFT + $U$). Based on the DFT + $U$ results, significant compositional deviations from stoichiometric LFO cannot be expected by providing rich or poor conditions of the oxidic precursor compounds (Fe$_2$O$_3$ and La$_2$O$_3$) in a solid-state processing route. In the experimental part, LFO was synthesized with a targeted La-site deficiency. We analyze the resulting phases in detail by X-ray diffraction and dedicated microscopy methods, namely, scanning electron microscopy and (scanning) transmission electron microscopy in combination with energy-dispersive X-ray spectroscopy and electron energy-loss spectrometry. Instead of a variation of the La/Fe ratio, a mixture of the two phases Fe$_2$O$_3$ and LFO was observed, resulting in an invariant charge state of Fe, which is in line with the theoretical results. We discuss our findings with respect to partly differing assumptions made in previously published studies on this material system.

# 1   Introduction

Owing to their ability to transform chemical energy into electrical energy and vice versa, solid-oxide fuel cells and solid-oxide electrolyzer cells (SOFCs and SOECs) are widely regarded as key technologies to realize the energy transition process [1]. For example, in an innovative approach ("Power-to-X"), SOECs are utilized to transform water and $CO_2$ into $H_2$ and CO by so-called co-electrolysis [2], [3], using renewable excess energy (from wind and sunlight) available in times of low demand. The produced gases can be stored for later use and then converted back into electrical energy by SOFCs [4], or further processed by the chemical industry [2]. In order to be cost-effective, SOFC/SOEC devices need to supply high power densities, which requires a high catalytic activity of the electrodes, that is fast rates of the oxygen exchange reactions at the surfaces and a high mobility of oxygen ions [5].

Perovskite compounds $ABO_{3-\delta}$, with large cations of fixed valence on the A sites ($La^{3+}$, $Ba^{2+}$, $Sr^{2+}$, and $Ca^{2+}$), transition-metal cations of mixed valence on the B sites ($Fe^{3+/4+}$, $Co^{3+/4+}$, and $Mn^{3+/4+}$), and an oxygen non-stoichiometry ($\delta$) have been widely studied as oxygen electrodes [6]–[9]. This is mainly due to the capability of the transition-metal ions to adopt different charge states (i.e., to form a redox couple) enabling the catalytic reactions, as well as due to the presence of oxygen vacancies in these compounds [10]–[12]. For example, in an SOEC, oxygen vacancies can incorporate the oxygen ions adsorbed at the surface, and mediate their transport to the interface with the electrolyte [5], [13].

Recently, defect engineering has become a strongly emerging discipline in the context of perovskites, which aims at improving their catalytic performance by tailoring the stoichiometry and by incorporating defects at specific concentrations [14], [15]. For example, due to charge compensation, the partial replacement of $La^{3+}$ by $Sr^{2+}$ in $La_{1-x}Sr_xFeO_3$ leads to an increased concentration of both, $Fe^{4+}$ ions and oxygen vacancies, which enhance the activity of the catalytic reaction substantially [16]. Similarly, the charge state of Mn can be actively manipulated by replacing $La^{3+}$ with $Ca^{2+}$ in $La_{1-x}Ca_xMnO_{3+\delta}$, where the oxygen excess does not correspond to interstitial sites, but is effectively realized by a complete oxygen lattice and cation vacancies [17].

Since A-site vacancies in perovskites are generally negatively charged, high concentrations of these point defects would affect the catalytic properties in a similar way as doping with lower-valent cations. In the scope of defect engineering, the synthesis of cation-deficient perovskites is therefore regarded as a promising optimization route [18]. In addition to oxidized transition-metal ions, the negative charges of A-site vacancies can in principle be compensated as well by positively charged oxygen vacancies, which are vital for the catalytic process.

Within the class of catalytically active $ABO_3$ perovskites, $LaFeO_3$ (LFO) is an important representative of a mixed ionic and electronic conductor. LFO can be regarded as a base, or parent material for electrodes in SOFC/SOEC devices, which is discussed extensively in a recent review article by Pidburtnyi et al. [19]. As compiled there, numerous studies analyze A- and B-site doping of this compound, that is $La_{1-x}A_xFe_{1-y}B_yO_{3-\delta}$, aiming at improving its catalytic properties.

In our study, we explore the suitability of LFO as a prototype compound of this material class for A-site vacancy engineering by a solid-state processing route. Applying a combined theoretical and experimental approach, we analyze defect and phase formation properties to emphasize the challenges and difficulties in this context. In particular, the question is whether a targeted La deficiency in the initial precursor substances before the synthesis process can lead to a corresponding amount of La vacancies in the final compounds and to the formation of desired $Fe^{4+}$ ions and O vacancies in sufficiently high concentrations, as reported in the literature [20]–[22]. To answer this question, the probabilities of the formation of these point defects under realistic processing conditions are analyzed, and the phases, which actually form in the synthesis process, are examined.

We derive formation energies of point defects in different charge states within the stability region of LFO in the phase diagram by means of first-principles calculations based on density functional theory with a Hubbard-U correction (DFT + $U$). Such formation energies for LFO were calculated before by several authors [23]–[25]. However, partly inconsistent values were reported. We analyze and discuss the differences among these, as well as compared to our results, in the Supporting Information. Using the calculated formation energies, we obtain approximate defect concentrations as function of temperature and pressure of the oxygen atmosphere by determining the Fermi level, which leads to a charge neutral (i.e. charge-compensated) system. The procedure also yields concentrations of free charge carriers, that is, electrons in the conduction band ($e$) and holes in the valence band ($h$), which correspond to $Fe^{2+}$ and $Fe^{4+}$ ions, respectively [25], [26]. We discuss our results with respect to those of a model, originally developed by Mizusaki et al. [26], [27] and refined and applied to LFO by Wærnhus et al. [28], in which point defect and charge-carrier concentrations are obtained by fitting rate constants of proposed defect equilibrium equations to experimentally measured electrical conductivity data.

From a theoretical point of view, the stoichiometry achievable in the synthesis of LFO crystals results from the calculated ranges of defect concentrations. In an experimental approach, processing routes can be set up for an intentional formation of phases with specific non-stoichiometry by providing an excess or shortage of respective precursor compounds.

Several authors investigated non-stoichiometric LFO with targeted compositions $La_{1-x}Fe_{1-y}O_{3-\delta}$ experimentally [20], [22], [29]–[34]. The reported La deficiencies (x) are supposed to be determined by the amounts of precursor phases provided in the different synthesis processes. Scafetta and May [29] report large La deficiencies in LFO thin films of up to 22 % (x = 0.22), which they assume to be realized by corresponding high La vacancy concentrations. The catalytic activity of LFO was analyzed by Zhu et al. [20] and Wu et al. [22] for x between 0 and 0.2, with optima at x = 0.05 and x = 0.1, respectively, which is stated to be partly due to the existence of $Fe^{4+}$ ions at the particle surfaces. Zhu et al. [20] report the synthesis of a La-deficient single LFO phase up to x = 0.1, before $Fe_2O_3$ forms as a secondary phase for larger values of x. Faye et al. [30] described the formation of $Fe_2O_3$ on the surface, as evidenced in O 1s XPS spectra for x ≥ 0.1. Wærnhus et al. [31] already detected iron oxide by transmission electron microscopy (TEM) for much smaller La deficiencies of x = 0.003. In both of these studies it is ruled out that La vacancies exist in amounts corresponding to the targeted deficiencies. This agrees to some extent with the findings of Delmastro, Spinicci, and coworkers [32], [33], who detected nano-sized grains with stoichiometric LFO cores and Fe–O surface configurations accounting for the La shortage up to x ≤ 0.3.

In the experimental part of this work, we applied electron microscopy to analyze composition and phase formation, as well as the ionic charge of Fe, in intended La-deficient LFO in detail. To this end, we synthesized LFO, both in the form of powders and sintered pellets, by applying a solid-state route, weighing in different ratios of $La_2O_3$ and $Fe_2O_3$ precursor oxides with the aim to achieve non-stoichiometric compounds. The resulting phases were analyzed by X-ray diffraction (XRD) and visualized by scanning electron microscopy. The La/Fe ratio was measured by energy dispersive X-ray spectroscopy (EDS) using a transmission electron microscope. Electron energy loss spectra (EELS) were recorded and analyzed with respect to charge state variations of Fe.

The paper is organized as follows: after a description of the modeling of the LFO phase and the defect calculation procedure, we report the results of defect formation energies and defect concentrations. The results of the experimental measurements are described in the following, subdivided into the analysis of LFO powders (with XRD, EDS, and EELS) and pellets [with scanning electron microscopy (SEM) and scanning transmission electron microscopy (STEM)]. We then discuss the results, focusing on links between the theoretical and experimental findings, and we compare our results with statements made in the literature about the investigated material system. Details about the DFT $+ U$ calculations and about the synthesis process as well as the applied microscopy techniques and procedures are described in a separate methods section at the end of the paper.

## 2 Phase and Defect Modeling

### 2.1 Supercell Model of the LFO Crystal Structure

At room temperature, LFO crystallizes in the orthorhombic perovskite structure with the space group $Pnma$ (#62) [35], [36]. For the calculations, the structure was set up in a $2 \times 1 \times 2$ representation of the primitive orthorhombic $Pnma$ cell, with a supercell containing 16 formula units (80 atoms) (*cf.* Figure 1). This cell size and geometry correspond to a supercell with dimensions $2\sqrt{2} \times 2 \times 2\sqrt{2}$ of a cubic perovskite. It was chosen as a reasonable compromise with respect to computational costs and weak interactions of the point defects with their periodic images since point defects are separated by at least two times the cubic perovskite lattice parameter. The chosen cell size ensures the correct (experimentally observed) orthorhombic space group symmetry of the crystal structure is preserved in the calculations with defects. La atoms are located at the Wyckoff position (WP) $4c$ at coordinates $(x, 0.25, z)$, Fe atoms at the WPs $4b$ $(0, 0, 0.5)$, and O atoms at the WPs $4c$ $(x, 0.25, z)$ (O1) and $8d$ $(x, y, z)$ (O2). Values for $x$, $y$ and $z$ for the different elemental sites are given in refs. [35], [36]. Periodic boundary conditions were applied for all the DFT $+ U$ calculations in this study. Computational details are given in the methods section. The structural relaxation of the positions of all atoms and of the volume of the supercell results in the lattice parameters (i.e. lengths of the primitive cell vectors) of 5.644 Å (in [100] direction with respect to the $Pnma$ convention), 7.911 Å ([010]) and 5.576 Å ([001]). These values are in reasonable agreement with the lattice parameters of LFO derived experimentally in this work by XRD measurements at room temperature (*cf.* Table 2).

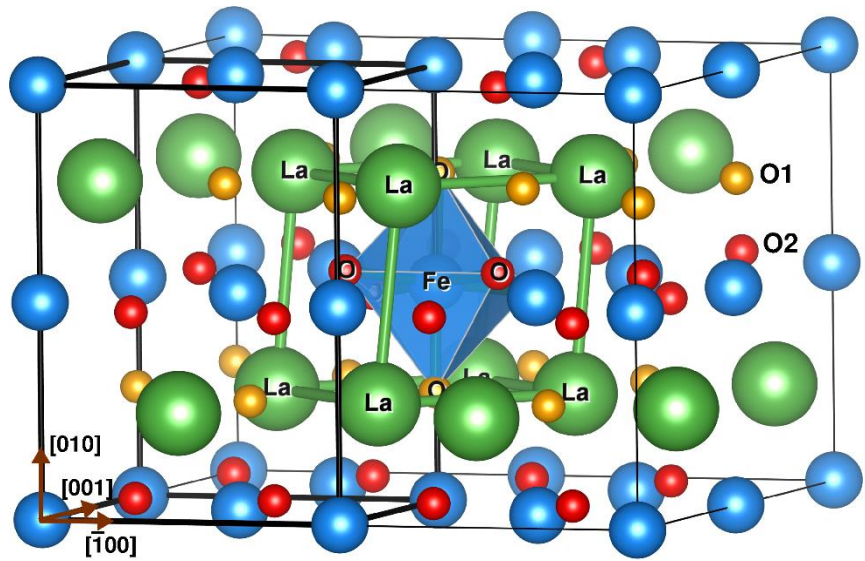

**Figure 1. Orthorhombic supercell of LFO consisting of 16 formula units (80 atoms). The primitive crystallographic cell ($Pnma$) is highlighted (bold black). Additionally, the typical ABO₃ perovskite motif of an octahedrally coordinated transition-metal cation (B position) in the center of a cube with cations at the corners (A position) is displayed (green). Compared to the cubic perovskite structure, this cube is slightly distorted in the orthorhombic structure, and there are two symmetry-inequivalent oxygen sites, which are distinguished by orange spheres (O1, in the La layer) and red spheres (O2, in the Fe layer).**

## 2.2 Calculation of Formation Energies and Concentrations of Point Defects

The formation energy $\Delta E_{i,q}$ of a point defect $i$ in a charge state $q$ was calculated using the well-established formula for the so-called supercell approach, which is valid in the dilute limit, that is for non-interacting defects [37]:

$$\Delta E_{i,q}(\{\Delta\mu_j\}, E_\mathrm{F}) = E_{i,q} - E_\mathrm{bulk} + \sum_j n_j \left(\mu_j^{(0)} + \Delta\mu_j\right) + q(\epsilon_\mathrm{VBM} + E_\mathrm{F}) + E_{i,q}^\mathrm{corr} \qquad (1)$$

$E_\mathrm{bulk}$ is the total energy of the defect-free bulk supercell, and $E_{i,q}$ is the total energy of a corresponding system containing one point defect $i$ in the charge state $q$. This defect can be imagined as being formed during the synthesis process of the material by the exchange of $n_j$ atoms of type $j$ between an elemental reservoir and the compound (e.g. $n_j < 0$ in the case of interstitials and $n_j > 0$ in the case of vacancies). The energy of such a reservoir—the chemical potential $\mu_j$—is generally referenced to the energy $\mu_j^{(0)}$ of the element $j$ in its ground-state structure. We define the chemical potentials referenced in this way as $\Delta\mu_j$, with $\Delta\mu_j = \mu_j - \mu_j^{(0)}$, and the set comprising the values of this quantity for all species $j$ involved in the formation of the defect is denoted by $\{\Delta\mu_j\}$. The sum runs over all of these species, which in our case can be La, Fe, and O, with the hexagonal (space group $P6_3/mmc$), body-centered cubic ($Im\bar{3}m$), and molecular (O$_2$) reference ground-state structures, respectively.

Figure 2 displays the ranges of values of $\Delta\mu_j$ for $j = $ La, Fe, and O, which lead to stable LFO, in a phase diagram (colored area). This extended stability region was derived from the formation energies of LFO and competing oxide phases containing La and/or Fe, as described and calculated in a previous study [38]. In the quantum-mechanical treatment of oxygen as an ideal gas, the chemical potential of oxygen can be expressed by the temperature and partial pressure of the oxygen atmosphere [39]. Since we use this relationship frequently throughout our work, we explicitly describe it in the Supporting Information.

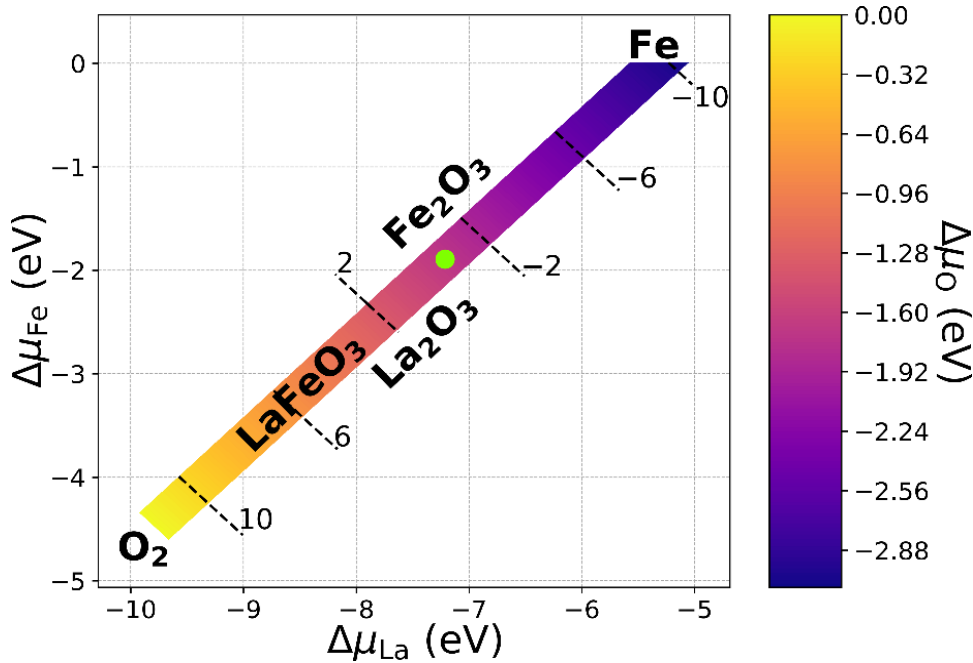

**Figure 2.** Phase diagram of LFO with respect to the chemical potentials of the elemental components. Crossing the borderlines of the stability region (colored area) leads to the formation of the denoted unary and binary phases in their ground states. For a given gas temperature $T$, $\Delta\mu_O$ can be expressed by the pressure $p$ of the oxygen atmosphere. For $T = 1400$ K, the pressure values are given at the dashed black lines in the logarithmic scale [$\log_{10}(p/p_0)$ with $p_0 = 1$ atm]. The light green circle marks the point in the phase diagram of which the formation energies are given in Table 1.

In addition to the exchange of atomic species with reservoirs, electrons are exchanged with an electron reservoir if charged defects are formed. The chemical potential of that reservoir is described as the sum of the valence band maximum ($\epsilon_{\text{VBM}}$, calculated for the perfect crystal) and $E_F$, which denotes the position of the Fermi level relative to the valence band maximum (VBM). An energy term $E_{i,q}^{\text{corr}}$ was taken into account following the procedure of Freysoldt et al. [40] in order to correct for the artificial Coulombic interaction between a charged defect and its own periodic images, as well as to align the different energy references of charged and neutral cells. In this work, we performed the charge corrections by applying the software *PyCDT* developed by Broberg and coworkers [41].

In the limit of dilute defects, the dependence of the concentration $x_{i,q}$ of a point defect $i$ in the charge state $q$ on the temperature $T$ of the system can be obtained from the defect formation energy $\Delta E_{i,q}$ via

$$x_{i,q} = c_i \Theta_{i,q} \exp\left(-\frac{\Delta E_{i,q}}{k_B T}\right) \qquad (2)$$

with the Boltzmann constant $k_B$. $c_i$ is the concentration of the lattice sites which can be occupied by the defect. The effect of the total entropy and volume change associated with the formation of a defect on its concentration is accumulated in the factor $\Theta_{i,q}$. In this study, the entropy resulting from equivalent electronic configurations due to the degeneracy of the band edges was taken into account in the way described by Ma *et al*. [42].

Using the concentrations of defects derived in the way described above, the Fermi level was determined self-consistently from the condition of overall charge neutrality [43]:

$$e + \sum_i \sum_{q<0} |q| x_{i,q} = h + \sum_i \sum_{q>0} q x_{i,q} \qquad (3)$$

$e$ and $h$ represent the concentrations of electrons in the conduction band and the concentrations of holes in the valence band, respectively. The values depend on the effective masses $m^*$ of the charge carriers at the respective band edges, which were derived by Singh *et al*. [44] for LFO from all-electron DFT calculations. Averaging these values over all high symmetry directions of the degenerate band edges [42] yields $m_e^* = 0.94 m_e$ and $m_h^* = 4.88 m_e$, with $m_e$ denoting the mass of a free electron. The band gap was set to 2.3 eV as it was derived by DFT + $U$ calculations of LFO with $U_{\text{Fe}} = 4$ eV in a previous study [38] and also in this work (*cf.* Figure 9). The value agrees with the band gap determined experimentally by Scafetta et al. [21].

## 3 Theoretical Results

### 3.1 Point Defect Formation Energies

Applying equation 1, we calculated the formation energies for the following point defects: La vacancies ($V_{\text{La}}$) and iron vacancies ($V_{\text{Fe}}$), both in the charge states [0, –1, –2, –3], O vacancies ($V_O$) in the charge states [0, +1, +2], and the antisites Fe on La ($\text{Fe}_{\text{La}}$) and La on Fe ($\text{La}_{\text{Fe}}$), both in the charge states [–1, 0, +1]. Oxygen vacancies were considered on the two inequivalent sites for oxygen atoms in the orthorhombic perovskite structure (*cf.* Figure 1). The charge states of the defects were chosen according to the number of valence electrons which the empty site or replaced ion can reasonably accept (−) or donate (+), based on the oxidation states that the ions adopt in the perfect crystal.

Table 1 lists the defect formation energies evaluated at a specific point in the phase diagram, namely $\Delta\mu_{La} = -7.21$ eV, $\Delta\mu_{Fe} = -1.89$ eV, and $\Delta\mu_{O} = -1.72$ eV, which is marked by the light green circle shown in Figure 2. This point is located at the center of the line corresponding to an oxygen pressure of 0.2 atm, which is the atmospheric partial pressure of oxygen, at a temperature of 1400 K, reasonably representing the synthesis conditions described in the methods section. To obtain the formation energies of the charged defects, we chose a Fermi level $E_F$ of 0.89 eV. As it will be shown in the following subsection, this value of $E_F$ leads to a charge neutral system at the specified chemical potentials. At this point, the dominant defects are $V_{La}^{-3}$, $V_{Fe}^{-3}$, $Fe_{La}^{-1}$, $Fe_{La}^{0}$ and $V_{O2}^{+2}$. The vacancies $V_{O1}$ have higher formation energies than the $V_{O2}$. While the distances to the two nearest Fe neighbors are almost identical for O1 and O2, the distance to the next La atom is by about 1.5 % shorter for O1. The smaller distance corresponds to a stronger bonding and therefore to a higher formation energy of the vacancy. It has to be noted that a variety of different values for point defect formation energies are reported for LFO in the literature [23]–[25], even though they were all obtained by DFT methods. We discuss these differences in the Supporting Information.

Table 1. Point Defect Formation Energies of LFO in Different Charge States $q$, Evaluated at $\Delta\mu_{La} = -7.21$ eV, $\Delta\mu_{Fe} = -1.89$ eV, and $\Delta\mu_{O} = -1.72$ eV, and at a Fermi Level $E_F = 0.89$ eV.

|        | $q = -3$ | $q = -2$ | $q = -1$ | $q = 0$ | $q = +1$ | $q = +2$ |
|--------|----------|----------|----------|---------|----------|----------|
| $V_{La}$ | 1.24 | 2.10 | 3.00 | 4.13 | — | — |
| $V_{Fe}$ | 1.69 | 2.02 | 2.75 | 3.55 | — | — |
| $V_{O1}$ | — | — | — | 3.20 | 2.53 | 2.07 |
| $V_{O2}$ | — | — | — | 3.11 | 2.50 | 1.91 |
| $Fe_{La}$ | — | — | 1.56 | 1.78 | 2.91 | — |
| $La_{Fe}$ | — | — | 4.66 | 3.38 | 3.98 | — |

Figure 3 displays the charge transition levels $\varepsilon(q/q')$ for the considered defects. At these values of the Fermi level, defects in the charge states $q$ and $q'$ have the same formation energy. For $E_F < \varepsilon(q/q')$, the defect favors to adopt the charge state $q$, while for $E_F > \varepsilon(q/q')$, the charge $q'$ is more likely. As expected, $V_{La}$ is a strong acceptor, adopting the charge $q = -3$ for almost all accessible Fermi levels in the band gap. $V_{Fe}$ is a weaker acceptor than $V_{La}$, since it can also be present in lower positive charge states for low values of $E_F$. The oxygen vacancies are deep donors, that is, they are neutral under n-type conditions and have the charge +2 under p-type conditions. $Fe_{La}$ acts as an acceptor for higher Fermi levels, and $La_{Fe}$ is charge neutral within a large portion of the band gap.

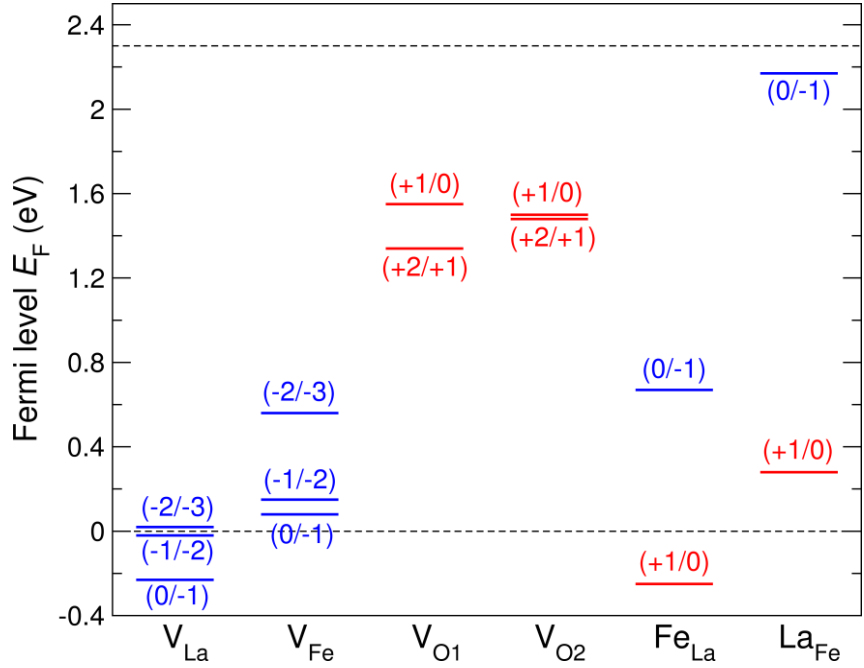

Figure 3. Charge transition levels $\varepsilon(q/q')$ of point defects in LFO. Acceptor levels (transitions between negative states) are marked in blue, and donor levels (transitions between positive states) are marked in red. The horizontal dashed lines mark the upper and lower borders of the band gap between the valence band edge (at 0 eV) and the conduction band edge (at 2.3 eV).

### 3.2 Defect and Charge Carrier Concentrations

Based on the formation energies of the defects in all considered charge states, and the corresponding defect concentrations derived by applying equation 2, the Fermi levels at which there is charge neutrality equation 3 were calculated for a temperature $T = 1400$ K at each point in the chemical potential space of stable LFO. The result is shown in Figure 4a as a function of the chemical potential of oxygen. In Figure 4b, the corresponding point defect concentrations are displayed, as well as the concentrations of free charge carriers ($e$ and $h$), in addition as a function of the oxygen gas pressure.

At each value of the oxygen pressure (or $\Delta\mu_O$, cf. Supporting Information), there are extended regions of values of the Fermi level and of the defect and charge-carrier concentrations. These regions originate from the changes in $\Delta\mu_{La}$ and $\Delta\mu_{Fe}$ along a line of constant $\Delta\mu_O$ as depicted in the phase diagram (Figure 2). At each value of $\Delta\mu_O$, the point of lowest $\Delta\mu_{La}$ and highest $\Delta\mu_{Fe}$ in the stability region of LFO corresponds to Fe$_2$O$_3$-rich and La$_2$O$_3$-poor synthesis conditions, and vice versa. As performed in this work (cf. methods section), and also described elsewhere (refs. [45], [46]), the processing of LFO can be achieved via a solid-state route by mechanical mixing of Fe$_2$O$_3$ and La$_2$O$_3$ powders under constant atmospheric conditions. Therefore, for comparing the results of our simulations to experiments, referring to conditions of the binary precursor compounds Fe$_2$O$_3$ and La$_2$O$_3$ in terms of *rich* and *poor* is more reasonable than referring to such conditions of the unary elements La and Fe.

At $T = 1400$ K, the pressures corresponding to the stability region of LFO range from about 10$^{-10}$ atm to about 10$^{12}$ atm (cf. Figure 2). The defects not shown in Figure 4 (such as $\text{La}_{\text{Fe}}$ or the neutral vacancies) exist in considerably lower concentrations and therefore have a negligible influence on both the Fermi level and the charge-carrier concentrations. In case of the oxygen vacancies, we only display the sum of the concentrations of $V_{O1}^{+2}$ and $V_{O2}^{+2}$.

Referred to the VBM, the Fermi level drops from about 60 % of the value of the band gap to about 10 % with increasing oxygen pressure between $10^{-10}$ atm and about $10^{12}$ atm. This behavior originates from decreasing chemical potentials of La and Fe with increasing oxygen pressures (*cf.* Figure 2), and it leads to growing concentrations of the La and Fe vacancies. Higher amounts of these always negatively charged defects are almost completely compensated by holes in the valence band ($h$), especially since the concentration of the dominant positive lattice defect ($V_O^{+2}$) naturally decreases with increasing gas pressure. Therefore, the necessary amount of positive charge carriers can only be provided by lowering the value of the Fermi level. For pressures below about 1 atm, the negatively charged antisite defect $\text{Fe}_{\text{La}}$ becomes relevant since its concentration reaches values comparable to those of the dominating vacancies. The implications of the defect concentrations on the experimentally accessible compositions will be discussed in the discussion section.

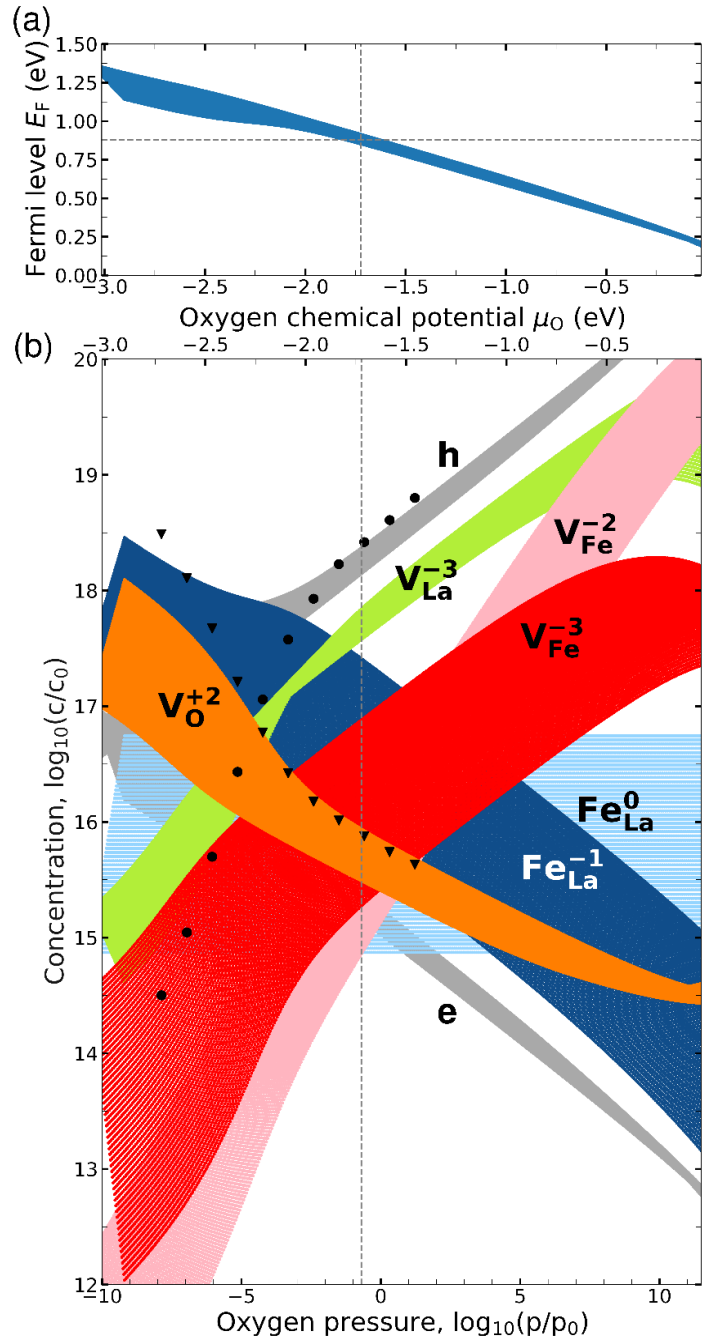

**Figure 4.** (a) Fermi level $E_F$, and (b) concentrations $c$ of dominant defects (extended colored regions) and free charge carriers ($e$, $h$; gray regions) on a logarithmic scale (with $c_0 = 1$ cm$^{-3}$) as a function of the oxygen chemical potential or, alternatively, of the corresponding oxygen gas pressure (logarithmic scale, with $p_0 = 1$ atm) at a temperature $T = 1400$ K. For the oxygen vacancies, the sum of the concentrations of $V_{O1}^{+2}$ and $V_{O2}^{+2}$ is displayed and denoted as $V_O^{+2}$. The vertical dashed line marks the atmospheric oxygen partial pressure (0.2 atm), as for example prevailing in the experimental synthesis of LFO in air in this work. The horizontal dashed line in (a) marks the equilibrium Fermi level at the specific chemical potential values marked by the green circle in Figure 2 and at the specified temperature. The black symbols shown in the lower graph (b) represent values reported by Wærnhus et al. [28] for the cation (circles) and oxygen (triangles) vacancy concentrations, derived at a temperature of 1273 K by fitting experimental results to model equations (*cf.* discussion section).

## 4 Experimental Results

In this section, we first present results of the crystallographic and compositional analysis of synthesized LFO powders with intended La deficiencies. Then, we describe the detailed phase analysis performed for sintered pellets. With the application of a higher temperature and higher mechanical load in the sintering process of the LFO pellets, we attempt to ensure no non-reacted precursor material remains which may influence the results.

### 4.1 LFO Powders

#### 4.1.1 Crystallography (XRD)

Figure 5 shows XRD diffraction patterns of the calcined powders with $La_2O_3/Fe_2O_3$ molar ratios $y = 1.0, 0.96, 0.94,$ and $0.9$. All of these samples have the orthorhombic perovskite structure ($Pnma$). On a four times enlarged y-scale in the right graph we plot the range from 31.5° to 36.5°, which contains the strongest LFO reflection peak 121, as well as the two strongest reflections 104 (at 33.118°) and $2\bar{1}0$ (at 35.612°) of $Fe_2O_3$. For the sample with $y = 0.9$, these two reflections are recognizable, as indicated by the arrows, and also for the two samples with $y = 0.94$ and $y = 0.96$, minor humps can be recognized at the same position, but with decreased intensity, which are however hardly above the noise. Table 2 lists the lattice parameters, as refined from Rietveld analysis of the full pattern of the main $Pnma$ phase. There is no significant dependence of the LFO lattice parameters on the $La_2O_3/Fe_2O_3$ molar ratio. The lattice parameters are in agreement with previously reported values by Falcón et al. [35] and Dixon et al. [36].

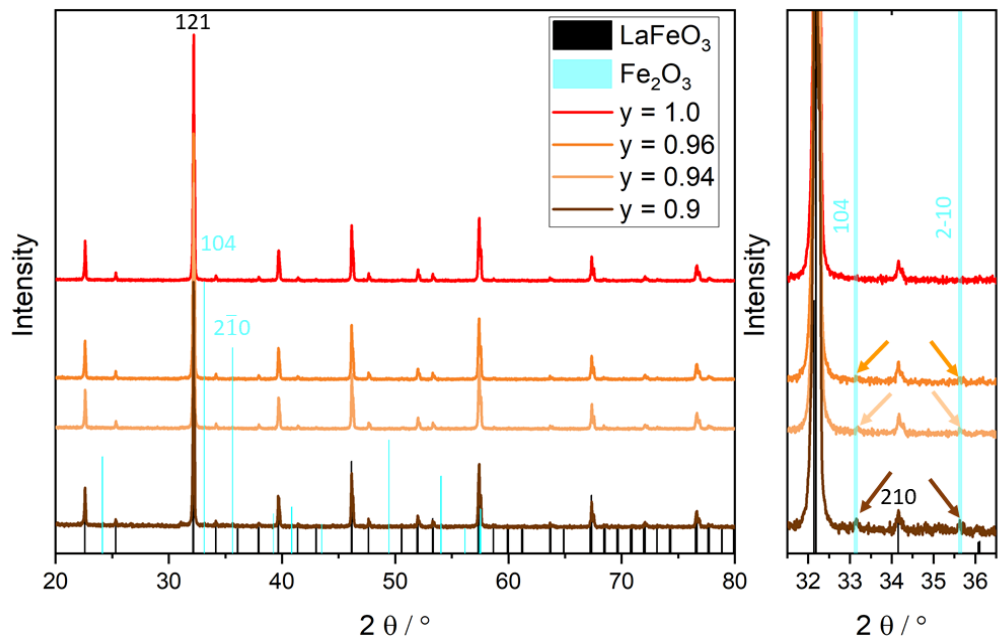

**Figure 5.** Left: XRD pattern of LFO with $La_2O_3/Fe_2O_3$ molar ratios $y = 1, 0.96, 0.94,$ and $0.9$. Right: Enlarged region for $2\theta$ between 31 and 37°, highlighting secondary phase reflections belonging to $Fe_2O_3$ for $y < 0.96$.

Table 2. Lattice Parameters of the $Pnma$ Perovskite Phase for Samples with $La_2O_3/Fe_2O_3$ Molar Ratios $y$ between 1 and 0.9, as Refined by the Rietveld Method from the XRD Patterns[a].

| $y$ | $a$ [Å] | $b$ [Å] | $c$ [Å] | $V$ [Å³] |
|---|---|---|---|---|
| 1.00 | 5.5659(1) | 7.8543(1) | 5.5548(1) | 60.7088(15) |
| 0.96 | 5.5647(1) | 7.8534(1) | 5.5546(1) | 60.6865(15) |
| 0.94 | 5.5649(1) | 7.8537(1) | 5.5544(1) | 60.6888(13) |
| 0.90 | 5.5653(1) | 7.8538(1) | 5.5544(1) | 60.6940(18) |

[a]The cell volumes V are given per formula unit.

### 4.1.2 Composition (TEM-EDS)

For the samples with La/Fe molar ratios $y = 1.00$, 0.96, 0.94, and 0.90, the La/Fe ratios as quantified from EDS in the TEM, are shown in Figure 6a. Spectra of 10 particles, mostly in the sizes range of 200–500 nm, as shown in Figure 6b, were acquired for each sample. By electron diffraction and image contrast it was checked that only perovskite particles were investigated by EDS. Quantification was performed with the $K_\alpha$ line (6.405 keV) for Fe and with the $L_\alpha$ line (4.647 keV) for La. The La/Fe ratio scatters around the value 1 for all samples, and no trend is visible. Similar results were obtained when the La K-line (La $K_\alpha$ = 33.442 keV) was used for quantification. Error bars are obtained from averaging over ten spectra (particles). Significant lanthanum deficiencies exceeding these errors cannot be detected. This indicates that LFO perovskite structures cannot be produced with a significant La deficiency by means of the employed processing route. Instead, iron oxide can form next to pure LFO, which is shown in Figure 6c for a large agglomerate of particles obtained for $y = 0.9$. Elemental maps of oxygen (red), iron (green), and lanthanum (blue) are overlaid on the high-angle annular dark-field (HAADF) image. From its yellow color, the particle at the lower right region of the agglomerate can be identified as iron oxide since blue (La) is missing. In the yellow region, elemental quantification results in a Fe/O ratio of 45:55, deviating from 40:60 for perfect $Fe_2O_3$, which can be attributed to absorption of the O $K_\alpha$ line as mentioned above. Such absorption effects are also responsible for the red/blue coloring of the larger part (pure LFO particles) of the agglomerate.

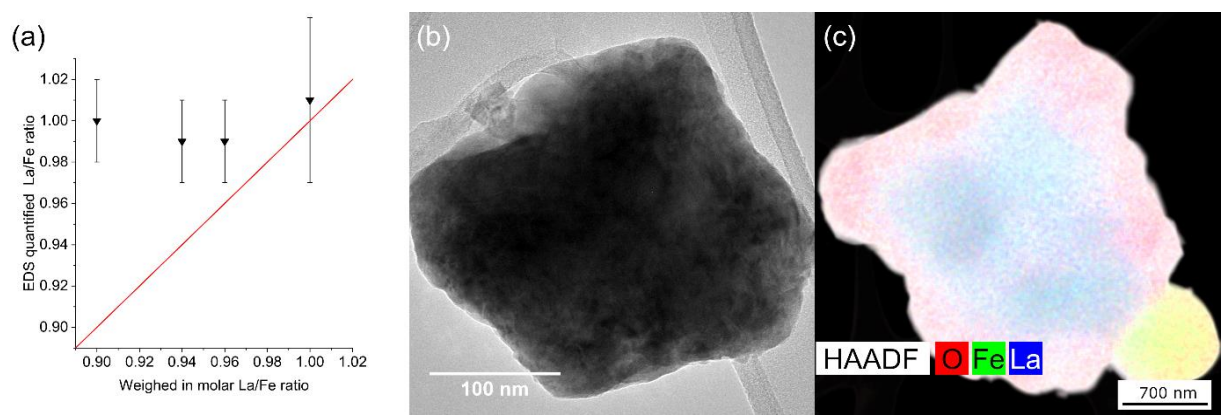

Figure 6. (a) La/Fe ratios measured by EDS (black triangles) for LFO samples with respect to the $La_2O_3/Fe_2O_3$ molar ratios $y = 1.0$, 0.96, 0.94, and 0.9 of the initial powder mixtures. The red line displays the values that would be expected if LFO phases were formed with La deficiencies according to the ratio $y$. (b) TEM image of a typical particle selected for the TEM–EDS analysis. (c) EDS maps, O (red), Fe (green), and La (blue), overlaid on the HAADF signal from a large agglomerate of several particles in the sample $y = 0.9$, showing an iron oxide particle at the bottom right.

### 4.1.3 Charge State of Fe (STEM–EELS)

To investigate the charge state of Fe ions inside the synthesized particles, we performed EELS in the STEM mode. In general, in EEL spectra, onsets and intensities (I) of the peaks, which belong to the ionization of core electrons, depend on the bonding environment and valence state of the atoms in the probed sample [47]. As references, we measured EEL spectra of iron oxide compounds with different oxidation states of Fe, namely, $Fe^{2+}TiO_3$ (ilmenite), $Fe^{2+}(Fe^{3+})_2O_4$ (magnetite), and $(Fe^{3+})_2O_3$ (hematite). The $L_{2/3}$ edges are displayed in Figure 7a. $L_2$ and $L_3$ denote the peaks arising from the excitation of electrons from the atomic states $2p^{j=1/2}$ and $2p^{j=3/2}$ in Fe, respectively, to unoccupied $3d$ states. In the spectra of the reference compounds, one can identify a chemical shift of the $L_{2/3}$ edge onset to higher energies, as well as an increasing intensity ratio $I(L_3)/I(L_2)$ with increasing Fe valence, which is in agreement with previous results of Tan et al. [48]. Figure 7b displays the Fe $L_{3/2}$ edge of the four powder samples of LFO with nominal $La_2O_3/Fe_2O_3$-ratios of 1.00, 0.96, 0.94, and 0.90, measured under similar conditions as the reference compounds. Neither a chemical shift nor a change in the peak intensity ratio $I(L_3)/I(L_2)$ is visible. We therefore exclude the presence of a significant amount of $Fe^{4+}$ ions, which would be expected in LFO structures with large concentrations of La vacancies.

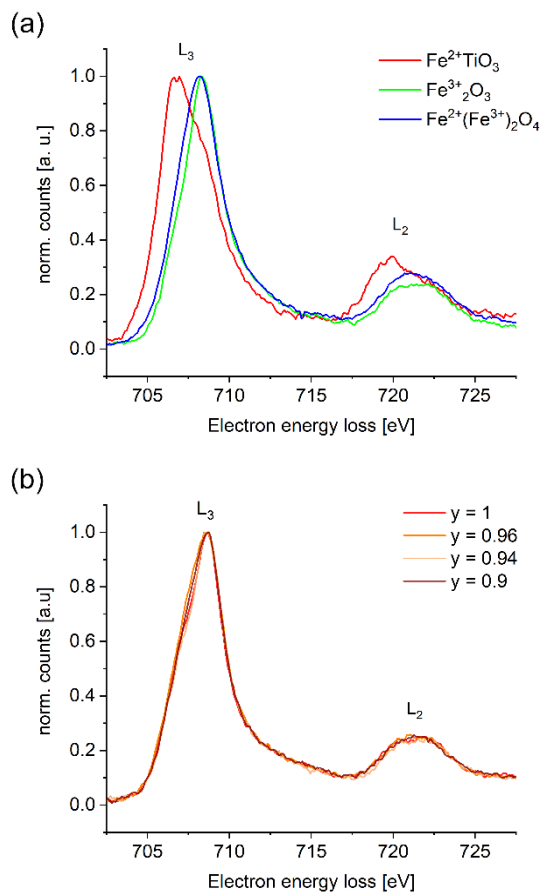

Figure 7. (a) EEL spectra of reference samples $Fe^{2+}TiO_3$, $Fe^{2+}(Fe^{3+})_2O_4$, and $(Fe^{3+})_2O_2$, (b) EEL spectra of samples with nominal $La_2O_3/Fe_2O_3$ molar ratios of $y = 1.0$, 0.96, 0.94, and 0.9.

## 4.2 LFO Pellets

### 4.2.1 Phase Analysis (SEM)

Pellets of samples with molar La/Fe ratios of $y = 1$, 0.96, and 0.9 were investigated by SEM. Figure 8a–c shows the obtained micrographs of back-scattered electrons (BSEs). In the stoichiometric LFO material ($y = 1$), the grain sizes range from 5 to 10 µm, and grains appear bright with slightly varying contrast depending on their orientation. The only other microstructural features are pores, which appear black. For the samples with $y < 1$, the grain size increases, porosity decreases, and a second phase with a medium gray level contrast appears. Exemplarily, two pores and two grains of the $Fe_2O_3$ secondary phase are marked by arrows. Quantification of an area of the size of 114,264 µm² (larger than the area displayed in Figures 8a–c) yielded an amount of $2.1 \pm 0.4$ % of secondary phase (corrected for porosity) for the sample with $y = 0.96$, and $4.6 \pm 0.6$ % for $y = 0.9$.

### 4.2.2 Atomic Structure (STEM)

From the pellet with $y = 0.96$, a self-supporting TEM sample was prepared and investigated in a probe-corrected STEM [49]. Figure 8d shows a $Fe_2O_3$ inclusion, confirming the SEM observation of the secondary phase. The contrast in the STEM-HAADF image is clearly darker since the La atoms with their high atomic number $Z$ are missing. Marked are the two regions from which additional high-resolution (HR) STEM micrographs were recorded. They are displayed in Figure 8e,f. The LFO structure shown in Figure 8e is oriented along [12−1] ($Pnma$) equivalent to the pseudocubic <201>$_{pc}$ orientation, showing clearly separated La planes (bright, $Z = 57$) and Fe planes ($Z = 26$) with less bright atoms. Structural models are overlaid for comparison. The $Fe_2O_3$ structure shown in Figure 8f is viewed along [42−1] ($R\bar{3}c$) with alternating distances between Fe columns that are present along this projection of the corundum structure. In this way, it is confirmed that $Fe_2O_3$ in the corundum structure (space group $R\bar{3}c$) is present as a secondary phase in the non-stoichiometric samples, and apparently, there is no other iron oxide.

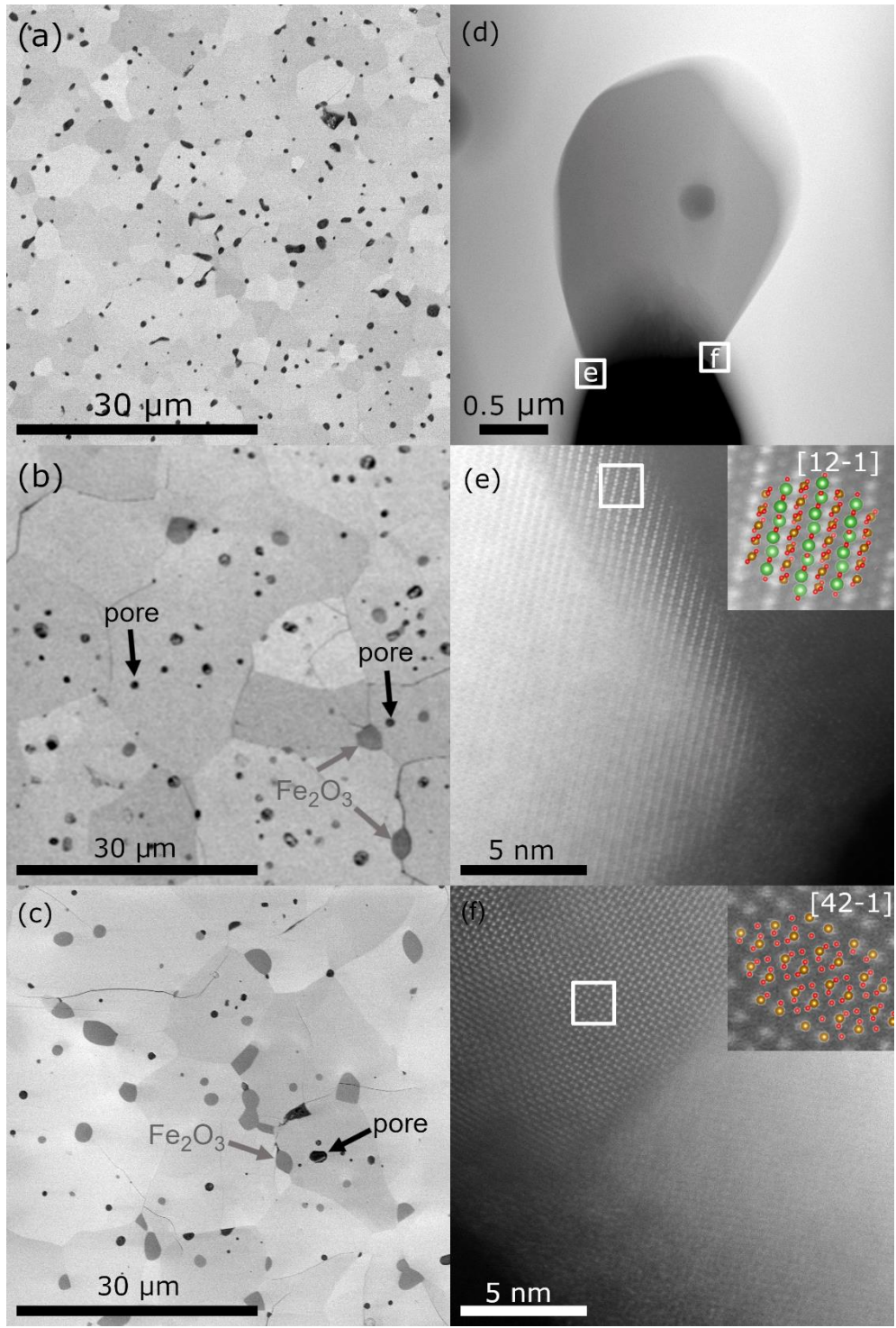

Figure 8. SEM BSE micrograph of (a) stoichiometric LFO (La/Fe molar ratio $y = 1$) showing only grains (bright) below 10 µm and pores (black); (b) Sample with $y = 0.96$: LFO grains (bright) are larger, porosity (dark) decreases, and a significant amount of the secondary phase $Fe_2O_3$ can be observed (medium gray, two grains are highlighted by arrows). (c) LFO and $Fe_2O_3$ grains for the sample with $y = 0.9$. (d) STEM-HAADF images of a $Fe_2O_3$ inclusion in the sintered pellet prepared from the sample with $y = 0.96$. (e) HRSTEM-HAADF micrograph at the left boundary of the inclusion showing the perovskite structure of LFO along $[12-1]_{Pnma}$ (<201>$_{pc}$) with separated Fe- and La-columns appearing with different intensities. The inset shows the magnified region with the projection of the LFO structure overlay, where Fe is displayed in brown, La in green, and O in red. (f) HRSTEM HAADF micrograph at the right boundary showing the $Fe_2O_3$ structure along $[42-1]_{R\bar{3}c}$ projection with alternating distances between columns of Fe ions (brown).

## 5 Discussion

In this section, we discuss the results on point-defect versus secondary-phase formation in LFO under La-deficient processing conditions. We analyze the results of our theoretical and experimental findings in mutual perspectives, and we refer and compare them to previously published reports on this topic.

From a thermodynamic point of view, point defects in crystal structures can exist in thermal equilibrium in a bulk material within certain concentration ranges. These ranges are determined by the values that the defect formation energies can have in the stability range of the material in a phase diagram with respect to open particle reservoirs. According to our calculations, at a temperature of 1400 K, the concentration of La vacancies in LFO in their highest ionization state (−3) increases from about $10^{15}$ to $10^{20}$ per cm$^3$ if the oxygen gas pressure changes between its lowest ($10^{-10}$ atm) and highest possible values (about $10^{12}$ atm) (*cf.* Figure 4). Outside of this pressure range, the formation of metallic Fe or molecular O$_2$, respectively, would be favored over perovskite LFO formation (*cf.* Figure 2). The concentration of oxygen vacancies in their preferred positive ionization state (+2) follows an opposite trend, decreasing from $10^{18}$ to about $10^{15}$ per cm$^3$ for increasing oxygen pressure. A qualitatively similar behavior was reported by Wærnhus et al. [28], as indicated in Figure 4 by the black symbols, which represent concentrations of oxygen vacancies (triangles) and cation vacancies (circles) given in their work. Lanthanum and iron vacancies cannot be distinguished by their applied method, which builds on a model originally developed by Mizusaki et al. [26]. The model delivers defect concentrations by fitting the equilibrium constants of mass-action laws of reactions between the point defects and atmospheric oxygen to experimental data [27], [50]–[52]. Quantitatively, the concentrations obtained by applying this model to LFO are up to 1 order of magnitude higher than our results, if we evaluate them at the same temperature as reported by Wærnhus et al. [28] (1273 K). According to equation 2, lower temperatures lead to lower point defect concentrations. The deviations can be caused by the existence of charge compensating defect complexes, such as Schottky defects $V_{La}^{-3} + V_{Fe}^{-3} + 3V_{O}^{+2}$, in the perovskite material [24], [28], for which the formation energies of the individual vacancies are lowered by a mutual electrostatic energy gain, compared to the case of dilute, that is, non-interacting charged point defects, as considered in our approach.

Based on our theoretical results (Figure 4), at atmospheric pressure and a temperature of 1400 K, the defect equilibrium mainly consists of $V_{La}^{-3}$, $Fe_{La}^{-1}$ and holes in the valence band (charge carriers $h$). Note that Wærnhus et al. [28] did not consider the antisite defect $Fe_{La}$ in their model, which can also contribute to some extent to the quantitative differences. However, in both models the amount of $V_{La}^{-3}$ does only negligibly influence the stoichiometry of LFO: with a total number of lattice sites in perfect LFO of around $8 \cdot 10^{22}$ cm$^{-3}$, a concentration of La vacancies of $10^{18}$ cm$^{-3}$ corresponds to La$_{0.99998}$FeO$_3$, and an increase in the number of La vacancies by an order of magnitude gives La$_{0.9998}$FeO$_3$. Applying rich or poor conditions of the precursor compounds (Fe$_2$O$_3$ and La$_2$O$_3$) in the synthesis has an even smaller influence since the concentration of La vacancies only changes by a factor of about two between these scenarios, which confine the extended data region of the La vacancies, as shown in Figure 4. Therefore, one cannot expect higher La deficiencies than 0.02 % in bulk perovskite LFO synthesized under the given processing conditions. Considering the change in the La vacancy concentration with oxygen pressure, there are no realistic conditions leading to stable LFO compounds with La deficiencies higher by several orders of magnitude (e.g. La$_{0.9}$FeO$_3$). This statement is valid for the bulk phase and basically independent of the processing method. However, wet-chemical synthesis routes from nitrates [20], [22], [33] can result in metastable compounds or in larger surface fractions compared to a solid-state processing. This may lead to different surface configurations at which locally a lower La stoichiometry may be realized.

This statement of our theoretical analysis is corroborated by our experimental analysis of the phases, which are formed in the synthesis of LFO with considerably higher intended La deficiencies of up to 10 %, as initially set by the molar La/Fe-ratios of the powder mixtures. In the XRD samples with intended La deficiencies of 4, 6, and 10 %, there are very small peaks at 33.112 and 35.612°, which can be attributed to the two strongest reflections 104 and $2\bar{1}0$ of Fe$_2$O$_3$ in the trigonal hematite (or corundum) structure with the space group $R\bar{3}c$. This is supported by the observation of a measurable amount of iron oxide, namely, Fe$_2$O$_3$, as a secondary phase in powder particles in the (scanning) transmission electron microscopy [(S)TEM], as well as in pellets in the SEM and STEM.

For all of the intended compositions of La$_y$FeO$_3$ with $y = 1$, 0.96, 0.94, and 0.9, quantification of the La/Fe ratio by EDS did not indicate La deficiencies in the single phase LFO particles, exceeding the experimental error of about 2 % (Figure 6a). Quantification of the amount of Fe$_2$O$_3$ in powder samples in the TEM is not possible, due to the small amount of the sample. For this purpose, we investigated polished cross sections of sintered pellets by SEM, which clearly revealed Fe$_2$O$_3$ as a secondary phase (Figure 8). For the two samples with nominal compositions $y = 0.96$ and $y = 0.9$, quantitative phase analysis yielded amounts of about $2.1 \pm 0.4$ and $4.6 \pm 0.6$ vol % of Fe$_2$O$_3$, respectively. From the densities of LFO (6.51 g/cm³) and Fe$_2$O$_3$ (5.03 g/cm³), one calculates 1.74 vol% for $y = 0.96$ and 4.52 vol% for $y = 0.9$ under the assumption that no La-deficient LFO can be formed. Therefore, our experimental results are in agreement with the conclusion made from our theoretical results, namely, that there is no considerable La deficiency realized by vacancies in the LFO structure. Our results are in line with experimental work of Wærnhus et al. [31], who detected both Fe$_2$O$_3$ and La$_2$O$_3$ secondary phases for both $y = 0.997$ and $y = 1.003$, respectively, corresponding to a narrow solid solution interval of LFO.

Considering the point defect concentrations (Figure 4), the formation of Fe$_2$O$_3$ can also be understood from another perspective. Independent of their charge states, the antisite defects $\text{Fe}_{\text{La}}$ have the highest concentrations within the accessible range of values at a given $\Delta\mu_\text{O}$ for the points in the phase diagram at the Fe$_2$O$_3$/LFO separation line, which correspond to La$_2$O$_3$ poor conditions (*cf.* Figure 2). An antisite defect $\text{Fe}_{\text{La}}$ within an LFO perovskite structure leads to the local stoichiometry of Fe$_2$O$_3$, which may act as a nucleation site for the formation of an Fe$_2$O$_3$ phase. This is more likely for higher total $\text{Fe}_{\text{La}}$ concentrations. At atmospheric oxygen pressure, a temperature of 1400 K, and La$_2$O$_3$-poor conditions, the concentration of $\text{Fe}_{\text{La}}$ reaches a value comparable to the concentration of the La vacancy $\text{V}_{\text{La}}$ and considerably exceeds it for lower pressures. Especially at grain boundaries, the energy barriers hindering the phase formation and growth are lowered with respect to their values in the bulk perovskite structure. As one can see in Figure 8b,c, the Fe$_2$O$_3$ phase mainly formed at boundaries, or in close vicinity to them, between differently oriented LFO grains.

In addition to the phase formation, we can also analyze the electronic structure of LFO from both theoretical and experimental points of view. The theoretical results (Figure 4) are an increasing concentration of p-type charge carriers $h$, that is, holes at the VBM, and a decreasing concentration of n-type charge carriers $e$, that is, electrons at the conduction band minimum, for an increasing oxygen pressure. This behavior was previously reported for pure [28], as well as for Ca- [51] and Sr-doped LFO [27]. Figure 9 shows the site- and orbital-projected electronic density of states of LFO, which we derived from our DFT + $U$ calculations. The VBM is composed of overlapping Fe 3d and O 2p states, which constitute the Fe–O bonds. Due to these bonds, Fe adopts the charge state +3 in LFO, and a hole in the VBM can be interpreted as a missing electron in this bonding structure, leading to a nominal charge state of Fe of +4. This is in agreement with theoretical results presented by Zhu et al. [25], who proposed p-type conductivity in LFO to occur via a hopping mechanism of hole polarons.

Based on the value of $h$, that is, the concentration of $Fe^{4+}$ ions, derived in this work for LFO with oxygen as an ideal molecular gas at a temperature of 1400 K and a pressure of 0.2 atm, the average charge state of Fe would be around 3.0001 at these conditions. This value is considerably below the sensitivity of the EELS experiments performed in this work to detect changes in charge states of Fe by analyzing shifts of the Fe $L_{3/2}$ edge (Figure 7). However, assuming La vacancy concentrations in LFO according to the intended deficiencies would lead to charge states of Fe between 3.12 for $La_{0.96}FeO_3$ and 3.3 for $La_{0.9}FeO_3$. According to the spectra of the reference materials $FeTiO_3$, $Fe_3O_4$ and $Fe_2O_3$, such charge state deviations would be visible in EELS. The result that no such changes are detected in EELS agrees with our findings of stoichiometric LFO discussed above.

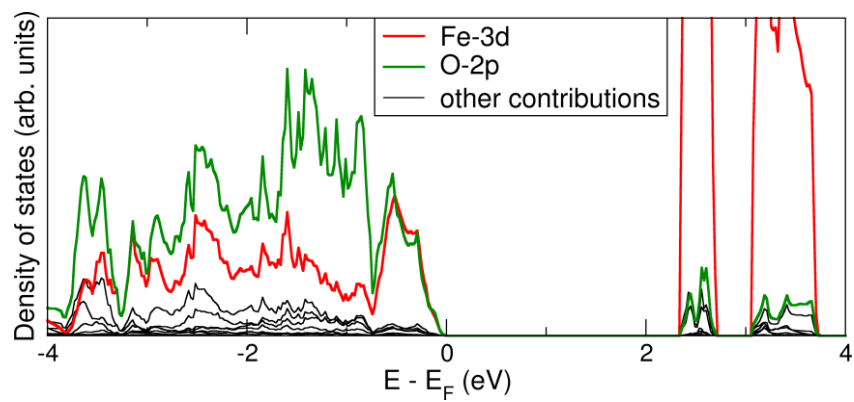

Figure 9. Site- and orbital-projected electronic densities of states of defect-free LFO.

Zhu et al. [20] report Fe charge states of 3.026, 3.034 and 3.024 for LFO with the targeted composition $La_yFeO_{3-\delta}$ and $y = 0.98$, 0.95 and 0.9, respectively. In contrast to our findings, they assume La deficiencies in the perovskite phases, which are realized almost completely by La vacancies. With the assumption of charge neutral systems, the reported charge states of Fe were then calculated from the nominal La deficiency and the oxygen deficiency, which was determined by iodometric titration. Scafetta and May [29] propose as well such a compensation effect between La and O vacancies for LFO with La deficiencies of up to 22 % because they have not observed a change in the charge state of Fe. However, for both of these studies (refs. [29] and [20]), the vacancy concentrations necessary to fulfill the stated LFO compositions exceed the values derived by considering defect formation energies in the phase diagram, as we have done in this study, by several orders of magnitude. In addition, due to the relationship between the chemical potentials of La and O (Figure 2), there are no accessible conditions, at which both of these values are simultaneously such that low vacancy formation energies and high concentrations of both, La and O vacancies would result. This is also reflected in the contrasting trend of these defect concentrations with changing oxygen pressure, as shown in Figure 4. According to thermodynamics, a stable perovskite structure of LFO cannot exist, which has vacancy concentrations in the bulk interior according to intended deficiencies of several percent.

According to this argument and in comparison to our experimental findings, we would expect the presence of binary iron oxide phases in the samples of Scafetta and May [29] and Zhu et al. [20]. XRD measurements were performed in both of these studies, and as it is apparent by the weak signals of the small amounts of the secondary phases we obtained by our XRD analysis in this work, the $Fe_2O_3$ peaks can easily remain undetected by this method as the signal-to-noise ratio can be reduced by fluorescence of Fe when Cu $K_\alpha$ radiation is used.

The strongest reflection of Fe$_2$O$_3$ is barely detectable for our samples with the nominal La/Fe-ratios y = 0.94 and y = 0.96 (Figure 5b). However, the reflection can be clearly identified for y = 0.9, and the intensity is systematically decreasing with increasing y, for which the presence of Fe$_2$O$_3$ was clearly confirmed by electron microscopy. Therefore, we are convinced that the small XRD bumps for y = 0.94 and y = 0.96 can be attributed to Fe$_2$O$_3$. Zhu et al. [20] interpret a slight shift of the strongest XRD reflection peak to higher angles as an indication for a shrinkage of the perovskite structure with increasing La deficiency, and they attribute this to the formation of La-vacancies and Fe$^{4+}$ ions. We have not observed such a variation of the lattice parameters (*cf.* Table 2). Delmastro et al. [32] state as well that a stoichiometry deviation in the bulk is to be excluded since the XRD peaks are at their angular positions, as calculated for LFO. Secondary phases can be missed by XRD. For La-deficient La$_{0.95}$Ni$_{0.6}$Fe$_{0.4}$O$_{3-\delta}$, Konysheva et al. [34] report very weak NiO reflections in the XRD pattern, but the refinement of the neutron powder diffraction data yielded a content of 5 mol % NiO, leading them to the conclusion that an A-site deficiency is not present.

Zhu et al. [20] and Wu et al. [22] employed surface-sensitive XPS as their spectroscopic method. In contrast to our EELS and EDS measurements performed in the (S)TEM, XPS measurements spatially average over all phases present in the sample. Hence, contributions of secondary phases such as Fe$_2$O$_3$ could be present without being detected. Faye et al. [30] showed by deconvolution of XPS O 1s spectra that a significant contribution of Fe$_2$O$_3$ is present already for $y = 0.9$. They attribute this to Fe$_2$O$_3$ formation at the surface and conclude that samples consist of stoichiometric LFO and Fe$_2$O$_3$ only. Therefore, in our opinion, evidence for the existence of La vacancies in bulk LFO in the high amounts stated by Scafetta and May [29] and Zhu et al. [20] cannot be deduced with certainty from their measurements. However, higher concentrations of La and/or O vacancies can indeed be present at particle surfaces, depending on the synthesis routes, where, due to different bonding environments, the corresponding defect formation energies can be lower than that in the bulk interior of the particles [22], [53]. This can lead to the formation of surface structures and stoichiometries differing from the bulk crystal, as for example reported by Delmastro, Spinicci, and coworkers [32], [33]. They detected nano-sized grains with stoichiometric LFO cores and FeO$_6$ octahedra at the surfaces for samples with targeted La deficiencies of up to 30 %.

## 6   Conclusions

In this work, we have studied non-stoichiometric LFO, both theoretically by atomistic DFT $+ U$ calculations and experimentally by XRD analysis, electron microscopy, and electron spectroscopy. We present calculated formation energies, charge transition levels, and concentrations of point defects, as well as concentrations of charge carriers and the Fermi level as a function of oxygen gas parameters in the solid-state synthesis process. Based on these results, the defect equilibrium ensuring a charge neutral system under conditions applied in this work, targeting at La-deficient LFO, mainly consists of La vacancies in their highest negative charge state (–3), holes in the valence band, which are effectively due to oxidized Fe ions (Fe$^{4+}$), and negatively charged Fe$_{La}$ antisite defects. To the best of our knowledge, the latter point defect type has not yet been considered in previously published studies dealing with defects in LFO. However, its existence in non-negligible amounts, especially under La$_2$O$_3$ poor synthesis conditions can facilitate the formation of iron oxide as a secondary phase.

By means of electron microscopy, we clearly detected and confirmed the presence of Fe$_2$O$_3$ in samples with targeted La deficiencies of up to 10 %, which were processed by a solid-state synthesis route. By EDS, we found agglomerates of Fe$_2$O$_3$ and LFO. Quantification of the La/Fe ratio in the perovskite particles did not hint at La deficiencies beyond the limits of the experimental errors of about $\Delta y = \pm 0.02$. In EEL spectra of the LFO particles, neither a chemical shift of the Fe $L_{2/3}$ edge nor a change in the $L_3/L_2$ intensity ratio was detected, both of which would be expected for structures with considerable amounts of La vacancies and, accordingly, $Fe^{4+}$ ions.

Our experimental findings are in line with our theoretical results. Even though La vacancies and $Fe^{4+}$ ions were identified to be the most favored defects at the specified conditions, their calculated concentrations in the range of $10^{17}$ to $10^{18}$ cm$^{-3}$ influence the stoichiometry only on the order of $10^{-5}$, which is way below the experimental sensitivity. Oxygen vacancies were found to have even lower concentrations in the bulk crystal of stable LFO under the applied conditions. This is at variance with previous studies, which assumed that off-stoichiometric single-phase LFO with high amounts of La vacancies can be synthesized, which are compensated by high amounts of oxygen vacancies to maintain a charge neutral system. Rather, the shortage of La in targeting La-deficient LFO leads to the formation of a two-phase composite material LFO/$Fe_2O_3$. We additionally confirmed this by analyzing SEM micrographs and HR STEM images of sintered LFO pellets.

The results emphasize that the method of defect engineering, especially for cation vacancies, faces the difficulties of thermodynamic stability and high vacancy formation energies under processing conditions realistic for solid-state synthesis. Our analysis of LFO serves as a case study, providing a starting point for further efforts to analyze phases and the defect structure of LFO-based $La_xA_{1-x}Fe_yB_{1-y}O_{3-\delta}$ compounds with a targeted off-stoichiometry. The combination of doping and vacancy engineering may be a promising route to find novel catalytically active materials with improved properties for SOFC/SOEC application.

## Methods

### Details of the DFT + $U$ Calculations

The ground-state structures and total energies of perfect and defective LFO crystals were obtained by DFT calculations using Vienna Ab Initio Simulation Package (VASP) [54]. A cutoff energy of 600 eV was used for the plane-waves basis describing the Bloch states of the valence electrons. The partial occupations of Bloch states were determined by the tetrahedron method with Blöchl corrections [55]. The interaction with the core electrons was treated by projector-augmented wave pseudopotentials [56]. The generalized gradient approximation in the formulation of Perdew et al. [57] was chosen as the exchange–correlation (xc) functional. Using approximate xc functionals, strongly correlated electrons, such as the 3d electrons in transition-metal oxides, turn out to be too delocalized due to uncompensated electronic self-interaction. In this work, this was corrected by applying a Hubbard $U$ correction [58] to the 3d electrons of iron (DFT + $U$). An $U$ value of 4 eV was applied, which was derived as an optimal choice in a previous study [38] by comparing experimental to theoretical oxidation energies of iron oxides [59]. Electronic self-consistency loops were stopped when the energy difference between two steps was less than $10^{-5}$ eV, and the structures were relaxed until the minimal force component acting on an atom was below 0.01 eV/Å. For the simulations of supercells containing point defects, the optimized cell parameters of the perfect crystal were fixed and only the internal coordinates were relaxed. The Brillouin zone integrations for the electronic structure optimization were performed on a Γ-point centered $k$-point mesh of dimensions $4 \times 6 \times 4$. The spins on the iron sites in LFO were arranged in antiferromagnetic order, such that the nearest neighbors of each spin are oriented in the opposite direction. This configuration is known to be favored in LFO below the Néel temperature of 750 K.

Material Synthesis

Lanthanum iron oxide materials were prepared by a solid-state reaction with $La_2O_3/Fe_2O_3$ molar ratios $y = 1.00$, 0.96, 0.94, and 0.90, aiming for stoichiometric or lanthanum-deficient $La_{1-x}FeO_3$ with $x = 0$ or $x = 0.04$, 0.06, and 0.1, respectively. The choice was made as comparable La deficiencies are reported in the literature [20], [22], [29], [30], [32], [33], and the step sizes should assure that corresponding changes in stoichiometry and Fe valence are detectable by EDS and EELS. Before weighing, the $La_2O_3$ powder (Alfa Aesar, purity 99.99%) was dried at 850 °C for 12h and then mixed with the appropriate amounts of $Fe_2O_3$ powder (Honeywell, purity > 99%). The materials were ball milled in isopropyl alcohol, dried in a rotation evaporator, calcined in air at 1100 °C for 8h, and ground with pestle and mortar to fine powders. These powders were analyzed by X-ray powder diffraction in Bragg–Brentano geometry (PANalytical Empyrean) using a Cu-$K_\alpha$ radiation source. The lower energy level for the radiation counted by the detector was adjusted to reduce fluorescence effects caused by Fe contained in the sample. Rietveld refinement of the lattice parameters was carried out with the program *FullProf* [60]. For samples with $La_2O_3/Fe_2O_3$ molar ratios $y = 1.00$, 0.96, and 0.90, we synthesized pellets by uniaxial die pressing and subsequent cold isostatic pressing, followed by sintering at 1350 °C for 8 h in air.

Electron Microscopy

To prepare the samples for the electron microscopy analysis, calcined powders of the samples with $La_2O_3/Fe_2O_3$ molar ratios $y = 1.00$, 0.96, 0.94, and 0.90 were dispersed in ethanol and applied to lacey carbon TEM grids. Both EDS and EELS experiments were conducted on powder samples using a Tecnai F20 electron microscope (FEI, USA [61]). For EDS, the TEM mode was chosen as the search for particles on the copper grid is facilitated, and spectra of single particles can be measured in few minutes by converging the beam on the selected particle. Quantification of the La/Fe ratio was performed with local TEM Imaging and Analysis (TIA) software according to the Cliff–Lorimer approach [62], which also includes background subtraction and signal integration [63]. EELS data were recorded in the STEM mode with a Gatan Tridiem 863P post column image filter (GIF). Additional EDS mappings of powder samples were performed using a Titan G2 80-200 Crewley (FEI, USA [49]), and spectrum images were analyzed by means of Esprit 1.9 software package by Bruker, USA.

For phase analysis with SEM, sintered pellets of $La_2O_3/Fe_2O_3$ molar ratios of $y = 1.00$, 0.96, and 0.90 were polished and investigated using BSE images recorded on a Quanta FEG 650 (FEI, Netherlands) operated at 5 kV with the Circular BackScatter detector. Phase analysis was performed using Olympus Stream Essentials Desktop version 1.9 and Count and Measure (Olympus, Japan) applying manual thresholds to separate bright LFO grains, dark pores, and grains of the $Fe_2O_3$ secondary phase, which is represented by intermediate gray levels. For STEM investigation, the sintered pellets were ground down to 100 nm and polished, subsequently dimpled with a model 656 dimple grinder, Gatan, USA, to about 20 nm thickness in the center, and finally ion milled (model 1051, TEM mill, Fishione, USA). HR STEM images of TEM samples prepared from sintered pellets were recorded on the Titan G2 80-200 Crewley, FEI, USA [49].

**Supporting Information**

Transformation formula between the chemical potential of oxygen and the oxygen gas temperature and pressure, comparison of defect formation energies with the literature

SI 1. The chemical potential of oxygen gas

The chemical potential of oxygen, $\mu_O$, is taken as one half of the chemical potential $\mu_{O_2}$ of the oxygen molecule, which can be expressed as the sum of $\mu_{O_2}^{(0)}$, the ground state energy of $O_2$ at zero temperature, and $\Delta\mu_{O_2}(T,p)$, the change of $\mu_{O_2}$ for finite temperatures $T$ and partial pressures $p$:

$$\mu_O(T,p) = \frac{1}{2}\mu_{O_2}(T,p) = \frac{1}{2}\mu_{O_2}^{(0)} + \frac{1}{2}\Delta\mu_{O_2}(T,p). \tag{A1}$$

We obtained a value for $\mu_{O_2}^{(0)}$ by calculating the total energy of an $O_2$ molecule by DFT, which we then corrected for the well-known over-binding error according to Ref. [38] and obtained $\mu_{O_2}^{(0)} = -8.57$ eV. For $\Delta\mu_O(T,p)$, we used the formula given by Samanta *et al.* [39], which can be derived by statistical quantum mechanics of an ideal gas [64]:

$$\Delta\mu_{O_2}(T,p) = k_B T \ln\left(\frac{p\lambda^3}{k_B T}\right) - k_B T \ln\left(\frac{Ik_B T}{\hbar^2}\right) + k_B T \ln\left[1 - \exp\left(-\frac{\hbar\omega_0}{k_B T}\right)\right] + \frac{1}{2}\hbar\omega_0, \tag{A2}$$

with the Boltzmann constant $k_B$ and the reduced Planck's constant $\hbar$. The translational, rotational, and vibrational degrees of freedom of a diatomic molecule are taken into account by the first, second and third term, respectively, on the right-hand side of Equation (A2). The corresponding characteristic quantities are the thermal de Broglie wavelength $\lambda = \hbar\sqrt{2\pi/mk_B T}$, the moment of inertia $I = m(d/2)^2$ and the vibrational frequency $\omega_0$ of the diatomic molecule, respectively. $m$ is the mass of the molecule and $d$ the bond length in the diatomic molecule. For $O_2$, $\omega_0 = 1568$ cm$^{-1}$ [39] and $d = 1.21$ Å. The last term on the right-hand side of Equation (A2) represents the quantum mechanical zero-point energy, which can also be formally attributed to $\mu_{O_2}^{(0)}$. However, with $\hbar\omega_0/2 \approx$ 0.015 eV, its influence on $\mu_O(T,p)$ is marginal.

Comparison of defect formation energies

Knowledge of the defect formation energies $\Delta E_{i,q}$ at specific values of $\{\Delta\mu_i\}$ and $E_F$ allows for the calculation of $\Delta E_{i,q}$ at other values $\{\Delta\mu_i'\}$ and $E_F'$ by applying Equation (1) (in the main paper):

$$\Delta E_{i,q}(\{\Delta\mu_i'\}, E_F') = \Delta E_{i,q}(\{\Delta\mu_i\}, E_F) + \sum_j n_j(\Delta\mu_j' - \Delta\mu_j) + q(E_F' - E_F). \tag{A3}$$

In Table 1 in the main paper, we list formation energies of point defects in LFO at $\Delta\mu_{La} = -7.21$ eV, $\Delta\mu_{Fe} = -1.89$ eV and $\Delta\mu_O = -1.72$ eV, and $E_F = 0.89$ eV. These formation energies were obtained by applying Equation (1) (in the main paper), with the total energies of the defective and perfect crystals calculated by DFT+$U$ with $U_{Fe} = 4$ eV. Formation energies of point defects in LFO have been reported previously [23]–[25], but in each case given at a different point in the phase diagram, and obtained using a different DFT approach. We used Equation (A3) to determine the literature values (Refs. [23]–[25]) of the formation energies of the neutral vacancies $V_{La}$, $V_{Fe}$ and $V_O$ at $\Delta\mu_{La} = -7.21$ eV, $\Delta\mu_{Fe} = -1.89$ eV, and $\Delta\mu_O = -1.72$ eV in order to compare these data to each other as well as to our results. The values are listed in Table SI1. Even without the additional complication of charge corrections ($q = 0$), the formation energies differ by up to 4 eV, which would lead to considerably different defect concentrations. We discuss possible reasons for this spread of data in the following.

Table SI1. Comparison of literature data [23]–[25] and our results for formation energies of the neutral vacancies $V_{La}$, $V_{Fe}$ and $V_O$ in LFO (in eV). Different values of $V_O$ correspond to $V_{O1}$ (upper value) and $V_{O2}$ (lower value). In addition, the band gaps reported in the literature are given. The $U$ value in the last line is given in parentheses, since there was no $U$ contribution considered in the HSE study of Zhu *et al*. [25], but based on results given in Ref. [38], a band gap of 3.5 eV can be obtained from DFT+$U$ calculations with $U_{Fe}$ = 8 eV.

|  | $U$ (eV) | $V_{La}$ | $V_{Fe}$ | $V_O$ | $E_{gap}$ (eV) |
|---|---|---|---|---|---|
| Ref. [23]; DFT | 0 | 2.32 | 2.70 | 2.49 | 0.8 |
| This work; DFT+$U$ | 4 | 4.13 | 3.55 | 3.20 3.11 | 2.3 |
| Ref. [24]; DFT+$U$ | 7 | 5.02 | 6.51 | 2.24 2.11 | 2.6 |
| Ref. [25]; DFT-HSE | (8) | 5.63 | 5.91 | 2.81 | 3.5 |

Pushpa *et al*. [23] employed DFT without a $U$ correction (equivalent to DFT+$U$ with $U_{Fe}$ = 0 eV), Taylor *et al*. performed DFT+$U$ with $U_{Fe}$ = 7 eV [24], and Zhu *et al*. applied DFT with a hybrid functional (HSE) [25]. The latter obtained a band gap $E_{gap}$ of LFO of 3.5 eV, which is considerably above the experimental value of 2.3 eV [21]. In a previous study, we found a linear relationship between $U_{Fe}$ and the band gap for LFO [38], which yields $U_{Fe} \approx 4$ eV for $E_{gap} = 2.3$ eV and $U_{Fe} \approx 8$ eV for $E_{gap} = 3.5$ eV. Therefore, we compare the DFT+$U$ values in Table SI1 with the HSE value of Zhu *et al*. [25] by interpreting the latter tentatively as resulting from DFT+$U$ with $U_{Fe}$ = 8 eV. In this regard, the formation energy of $V_{La}$ consistently increases if $U_{Fe}$ changes from 0 to 8 eV.

The formation energy of $V_{Fe}$ shows an increasing tendency with $U_{Fe}$, too. However, we have some doubt about the value obtained by Taylor *et al*. [24] for $U_{Fe}$ = 7 eV. As mentioned in a more detailed work of this topic by the same author (Ref. [65]), next to the total energies of LFO, the ground state energy of metallic iron, $\mu_{Fe}^{(0)}$, which enters Equation (1) (in the main paper), was calculated in Refs. [24] and [65] *also* within the DFT+$U$ approach with $U_{Fe}$ = 7 eV. In our opinion, this approach is questionable, since the properties of metals (good conductors) are usually better described by DFT *without* a $U$ correction (or with only small U values). To clarify this aspect, we performed a test calculation and obtained a lattice parameter of bcc iron of 3.03 Å for DFT+$U$ with $U_{Fe}$ = 7 eV compared to 2.83 Å for DFT without $U$. The latter agrees much better with the experimental value at room temperature of 2.87 Å. Furthermore, applying the DFT+$U$ value ($U_{Fe}$ = 7 eV) of $\mu_{Fe}^{(0)}$ as a test case, we obtain a formation energy of $V_{Fe}$ of 6.37 eV, which is close to the result reported by Taylor *et al*. (6.51 eV) [24]. In contrast, in our approach of the present work, we used the DFT value for $\mu_{Fe}^{(0)}$ (without a $U$ correction), and DFT+$U$ values with $U_{Fe}$ = 4 eV for the total energies of the defective and perfect LFO structures in Equation (1) (in the main paper). This is consistent to the calculation of formation energies of transition metal oxide compounds by combining DFT and DFT+$U$ values, and we corrected the systematic error introduced by this procedure as discussed in detail in Refs. [38], [66].

In the case of $V_O$, a systematic change of the formation energy with $U_{Fe}$ cannot be identified when comparing the literature values to our results (Table SI1). The over-binding of the $O_2$ molecule, which serves as the reference state for the calculation of the formation energy of the oxygen vacancy, is a well-known systematic inaccuracy of common DFT (LDA or GGA) calculations [59], [67]. We corrected this error in the calculation of $\mu_O^{(0)}$ by 0.64 eV, as derived in Ref. [38]. The necessity of taking into account such a correction was described in detail by Lee *et al*. [53]. The value they derived for the formation energy of $V_O$ in LFO agrees within 0.2 eV with our value. However, Refs. [23]–[25] do not report that the over-binding was taken into account in their calculation of $\mu_O^{(0)}$, which presumably is the reason for the consistently lower formation energies of $V_O$ obtained in these studies.

To sum up this discussion, the choice of the *U* value and the application of energy correction schemes considerably influences the point defect formation energies. In our opinion, the most reliable values can be obtained by applying DFT+*U* with $U_{Fe}$ = 4 eV for deriving the total energies of the LFO supercells, and by using DFT without *U* corrections for the total energies of the elemental ground state structures. In addition, one needs to correct the systematic errors originating from mixing DFT+*U* and DFT energies [66], as well as from the over-binding of the diatomic oxygen molecule [59]. $U_{Fe}$ = 4 eV yields the correct band gap for LFO [38] and is well established for a variety of iron-containing oxides. For this material class, including LFO, it was previously shown that using $U_{Fe}$ = 4 eV and applying the above compiled correction schemes also leads to compound formation energies in good agreement to available experimental data [38].

In contrast to the formation energies, the charge transition levels $\varepsilon$(q/q') reported in the literature [23]–[25] agree qualitatively well among each other and with the values obtained in this work (Figure 3 in the main paper). There is clear consistency that $V_{La}$ acts as a strong acceptor, that $V_{Fe}$ has a shallow acceptor level in the band gap, and that $V_O$ is a deep donor. However, together with the band gap, the levels shift to higher values for increasing $U_{Fe}$, as expected.

**AUTHOR INFORMATION**

**Corresponding Author:**

**Daniel Mutter** – Fraunhofer IWM, Wöhlerstraße 11, 79108 Freiburg, Germany; **ORCID:** https://orcid.org/0000-0003-3547-695X

**Corresponding Author:**

**Roland Schierholz** – Forschungszentrum Jülich GmbH, Institute of Energy and Climate Research (IEK-9), Wilhelm-Johnen-Straße, DE-52425 Jülich, Germany; **ORCID:** https://orcid.org/0000-0002-2298-4405

**Authors:**

**Daniel F. Urban** – Fraunhofer IWM, Wöhlerstraße 11, 79108 Freiburg, Germany; **ORCID:** https://orcid.org/0000-0001-8302-3504

**Sabrina A. Heuer** – Forschungszentrum Jülich GmbH, Institute of Energy and Climate Research (IEK-9), Wilhelm-Johnen-Straße, DE-52425 Jülich, Germany; RWTH Aachen University, Institute of Physical Chemistry, Landoltweg 2, DE-52074 Aachen, Germany; **ORCID:** https://orcid.org/0000-0002-9476-6909

**Thorsten Ohlerth** – Forschungszentrum Jülich GmbH, Institute of Energy and Climate Research (IEK-9), Wilhelm-Johnen-Straße, DE-52425 Jülich, Germany; RWTH Aachen University, Institute of Physical Chemistry, Landoltweg 2, DE-52074 Aachen, Germany; **ORCID:** https://orcid.org/0000-0002-2169-8765

**Hans Kungl** – Forschungszentrum Jülich GmbH, Institute of Energy and Climate Research (IEK-9), Wilhelm-Johnen-Straße, DE-52425 Jülich, Germany; **ORCID:** https://orcid.org/0000-0003-3142-3906

**Christian Elsässer** – Fraunhofer IWM, Wöhlerstraße 11, 79108 Freiburg, Germany; Freiburg Materials Research Center (FMF), University of Freiburg, Stefan-Meier-Straße 21, 79104 Freiburg, Germany

**Rüdiger-A. Eichel** – Forschungszentrum Jülich GmbH, Institute of Energy and Climate Research (IEK-9), Wilhelm-Johnen-Straße, DE-52425 Jülich, Germany; RWTH Aachen University, Institute of Physical Chemistry, Landoltweg 2, DE-52074 Aachen, Germany; **ORCID:** https://orcid.org/0000-0002-0013-6325

**ACKNOWLEDGEMENTS**

This research was funded by the German Federal Ministry of Education and Research (BMBF) in the framework of the Kopernikus project *Power-2-X*, grant numbers 03SFK2I0 (C.E., Freiburg) and 03SFK2Z0 (R.E., Jülich). The calculations were performed on the computational resource ForHLR I of the Steinbuch Centre of Computing (SCC) at the Karlsruhe Institute of Technology (KIT), funded by the Ministry of Science, Research, and Arts Baden-Württemberg and by the German Research Foundation (DFG). Crystal structures are visualized with VESTA [68].

2020, doi: 10.3390/ma13235555.

**Graphical Table of Contents**

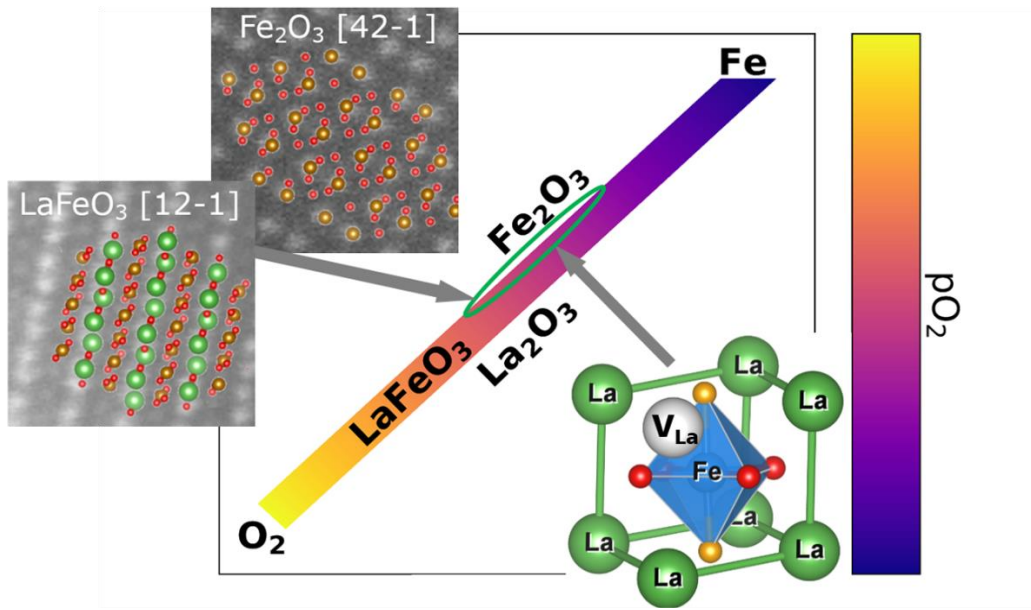